\documentclass[reprint]{JASA-EL}

\begin{document}

%% Version in square brackets will appear at top right of all pages except title page:
\title[JASA-EL/Sample JASA-EL Article]{End-to-End Audio-Visual Learning for Cochlear Implant Sound Coding Simulations in Noisy Environments}

\author{Meng-Ping Lin}
\altaffiliation{These authors contributed equally to this work.}
\email{mplin@media.ee.ntu.edu.tw}
\affiliation{Graduate Institute of Electronics Engineering
and the Department of Electrical Engineering, National Taiwan University, Taipei 106319, Taiwan}

\author{Enoch Hsin-Ho Huang}
\altaffiliation{These authors contributed equally to this work.}
\email{enoch.huang@citi.sinica.edu.tw}
\affiliation{Research Center for Information Technology Innovation Technology, Academia Sinica, Taipei 115201, Taiwan}

\author{Shao-Yi Chien}
\email{sychien@media.ee.ntu.edu.tw}
\affiliation{Graduate Institute of Electronics Engineering
and the Department of Electrical Engineering, National Taiwan University, Taipei 106319, Taiwan}

\author{Yu Tsao}
\email{yu.tsao@citi.sinica.edu.tw}
\correspondingauthor
\affiliation{Research Center for Information Technology Innovation Technology, Academia Sinica, Taipei 115201, Taiwan}
\affiliation{Department of Electrical Engineering, Chung Yuan Christian University, Taoyuan 320314, Taiwan}

\date{Published Online: 9 January 2026} 
\preprint{Author, JASA-EL}  %if you want want this message to appear in upper right corner of title page

\begin{abstract}
The cochlear implant (CI) is a successful biomedical device that enables individuals with severe-to-profound hearing loss to perceive sound through electrical stimulation, yet listening in noise remains challenging. Recent deep learning advances offer promising potential for CI sound coding by integrating visual cues. In this study, an audio-visual speech enhancement (AVSE) module is integrated with the ElectrodeNet-CS (ECS) model to form the end-to-end CI system, AVSE-ECS. Simulations show that the AVSE-ECS system with joint training achieves high objective speech intelligibility and improves the signal-to-error ratio (SER) by 7.4666 dB compared to the advanced combination encoder (ACE) strategy. These findings underscore the potential of AVSE-based CI sound coding. \end{abstract}

%% pacs numbers not used

\maketitle

%  End of title page for Preprint option --------------------------------- %

\section{Introduction}
\label{sec:introduction}
The cochlear implant (CI) is a hearing device designed to help individuals with severe-to-profound hearing loss regain partial hearing \cite{CI_review_zeng2022}. The CI system involves an external, upgradable sound processor that uses radio frequency to wirelessly link to a surgically implanted internal device. The CI system transforms speech signals into electrical pulse patterns, which are transmitted via the auditory nervous system to the auditory cortex of the brain, enabling sound perception. Developed through multidisciplinary collaboration, modern CIs now provide impressive speech recognition performance. With well-designed sound coding strategies that translate acoustic signals into electrode stimulation patterns, such as the widely used advanced combination encoder (ACE) strategy \cite{ACE_original_n_of_m_vandali2000}, most CI users can understand more than 80–90\% of speech sentences in quiet conditions \cite{CI_review_zeng2008}. However, further improvements are still needed, particularly in the processing of noisy speech.

Speech enhancement (SE) provides a potential pathway for improving CI in noisy environments. With the rapid development of deep learning, SE approaches for CIs are gradually shifting from conventional signal processing algorithms to data-driven neural networks \cite{BioASP_DDAE_lai2017, NR_CI_henry2023, DCCTN_mamun2024}.
Many SE approaches have been applied to CIs, including recent end-to-end architectures \cite{DeepACE_gajecki2023, AdversarialDeepACE_gajecki2025} that integrate both the pre-processing stage and the sound coding strategy \cite{strategy_review_wouters2015, NCU_CI_TNSRE_huang2021}, the core component responsible for converting audio into electrical stimuli.
However, these audio-only SE (ASE) approaches, which rely solely on audio input, do not make use of the visual cues available for processing background noise and overlapping speech. In general audio applications, audio-visual speech enhancement (AVSE), a common multimodal signal processing approach, has been proposed to incorporate visual information such as lip movements and facial expressions alongside audio signals, and has proven effective in outperforming audio-only speech enhancement methods \cite{AVSE_light_chuang2020, AVSE_light_chuang2022, AVSE1, AVSE2, AVSE3, AVSE4, AVSE5}.  Therefore, the exploration of visual cues and AVSE models for CI sound processing in noisy environments remains an important area of ongoing research \cite{BioASP_visualcues_tseng2021, BioASP_AVSE_lai2025}.

%Required for reprint format
\makeatletter
\@ifclasswith{JASA-EL}{reprint}{\newpage}{}
\makeatother

A potential way to improve the CI system using an AVSE module is to integrate it with the ElectrodeNet-CS (ECS) model \cite{ElectrodeNet_huang2024}.
 The core signal processing functions of the ACE coding strategy, envelope detection and channel selection (CS), are not differentiable, making them difficult to integrate with neural-network-based AVSE modules and refine through data-driven regression. To overcome this limitation, the ECS model replicates the core signal processing of ACE using a deep neural network (DNN) with a dedicated CS function, achieving comparable performance. Furthermore, the ECS coding strategy enables integration with diverse front-end and post-processing components, as well as joint training and end-to-end optimization. Consequently, the end-to-end AVSE-ECS CI system, which incorporates the AVSE pre-processing module and the ECS model, is proposed and validated using objective evaluations in this study.

This paper is organized as follows. Section~\ref{sec:method} explains the proposed AVSE-based CI system. Section~\ref{sec:experiments} describes the experimental setup for objective evaluation. Section~\ref{sec:results} offers the results and discussion. Section~\ref{sec:conclusion} draws the conclusion.

\section{Method}
\label{sec:method}

 \subsection{System Architecture}
\begin{figure}[tb]
	\centering
    \includegraphics[scale=0.65]{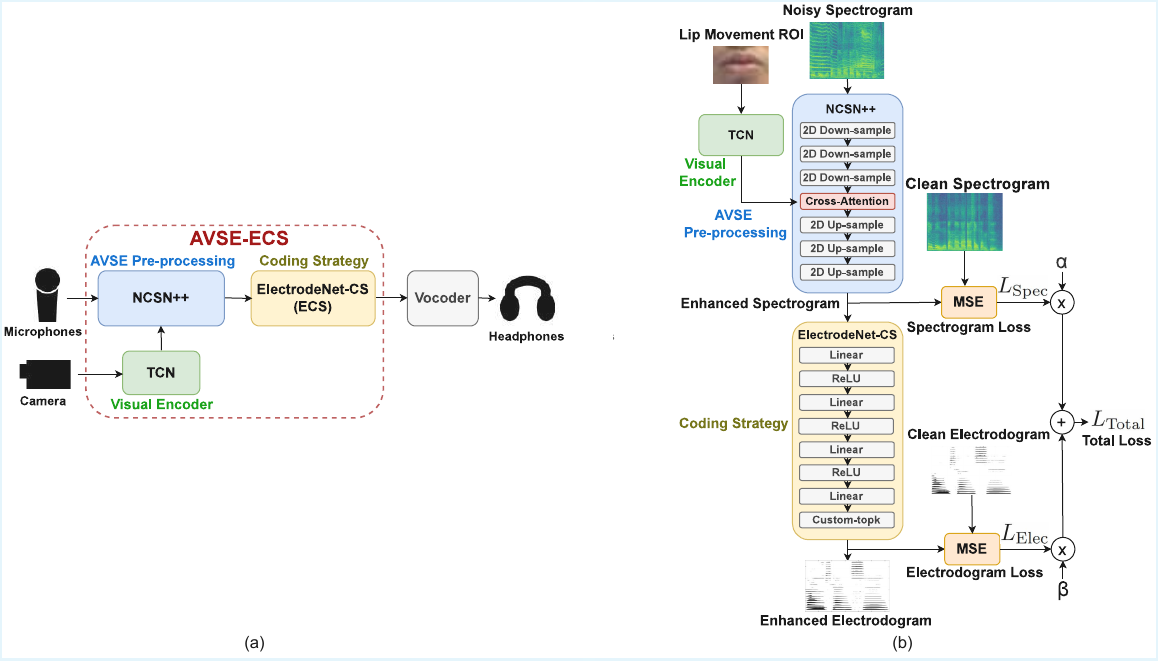}
    \caption{(a) The architecture of the proposed cochlear implant (CI) system, and the area enclosed by the dashed rectangle indicates the AVSE-ECS network. (b) Overall joint-training architecture for the AVSE-ECS network.}
    \label{fig:AVSE_ECS_framework}
\end{figure}

The system architecture of the proposed AVSE-ECS CI system is illustrated in Fig. \ref{fig:AVSE_ECS_framework}(a). The AVSE model incorporates auxiliary visual cues to generate enhanced speech, which is further processed by the ECS strategy into electrode stimulation patterns. A tone vocoder \cite{vocoder_tone_dorman2002} is used to convert electrode stimulation patterns into audible speech, which can then be analyzed using objective evaluation methods. Furthermore, this study introduces a joint training approach, and the architecture for training the AVSE-ECS network is illustrated in Fig. \ref{fig:AVSE_ECS_framework}(b). In this study, three CI network implementations are investigated: ECS (Fig. \ref{fig:Arch_comp}(a)), ASE-ECS (Fig. \ref{fig:Arch_comp}(b)), and AVSE-ECS (Fig. \ref{fig:Arch_comp}(c)).

\subsection{Coding Strategy}
This study adopts the DNN-based ECS model proposed in \cite{ElectrodeNet_huang2024} as the coding strategy, which is the core sound processing of the CI system. This model consists of four dense layers with 1024, 512, 256, and 22 neurons, followed by a custom topk layer designed for the CS function. The ECS model is pre-trained before being integrated into the AVSE-ECS system. Pre-training involved paired spectral–temporal matrices across L = 65 bins and M = 22 channels derived from clean speech processed with the ACE strategy. The training was configured to preserve key information in the $N_{topk} = 8$ channels selected by the network, while distributing lower envelope data across the remaining $M - N_{topk}$ channels, using the L1 norm across all M channels.

\subsection{Visual Encoder}
The temporal convolution network (TCN) is employed as the visual encoder as illustrated in Fig. \ref{fig:AVSE_ECS_framework}. This model consists of a 3D convolutional layer to process temporal sequences of frames, followed by a 2D ResNet-18. The extracted visual features are then passed to the AVSE model as visual embeddings. Note that a modified ResNet-18 visual encoder is pre-trained using the Lip Reading in the Wild (LRW) dataset \cite{LRW}. To provide the most relevant visual cues for SE, the encoder focuses on the mouth, referred to as the region of interest (ROI), rather than utilizing the entire facial region. Facial landmarks are detected using the Mediapipe face tracker \cite{Mediapipe}, and the mouth ROI is extracted accordingly. To reduce computational overhead during training, the visual encoder is kept frozen and the embeddings are pre-extracted, minimizing both memory usage and the number of training parameters.

\subsection{AVSE Pre-processing}
For the AVSE pre-processing, this study adopts and modifies the NCSN++ model from \cite{ScoreDiff}. As this original UNet is designed for generative tasks, the timestep is set to 1 in this study, and the model's output directly serves as the enhanced speech. For audio feature preprocessing before AVSE, the experimental settings are similar to \cite{CMKT}. Audio signals are resampled to 16 kHz and converted into complex-valued STFT representations. Using a window size of 510 and a periodic Hann window, the resulting frequency dimension is F = 256. The hop length is set to 128. Each STFT spectrogram is truncated to T = 256 time frames, enabling random cropping during batch training.

\subsection{Audio-Visual Fusion}
\begin{figure}[tb]
	\centering
    \includegraphics[trim={0 0 0 0.6cm},clip, scale=0.65]{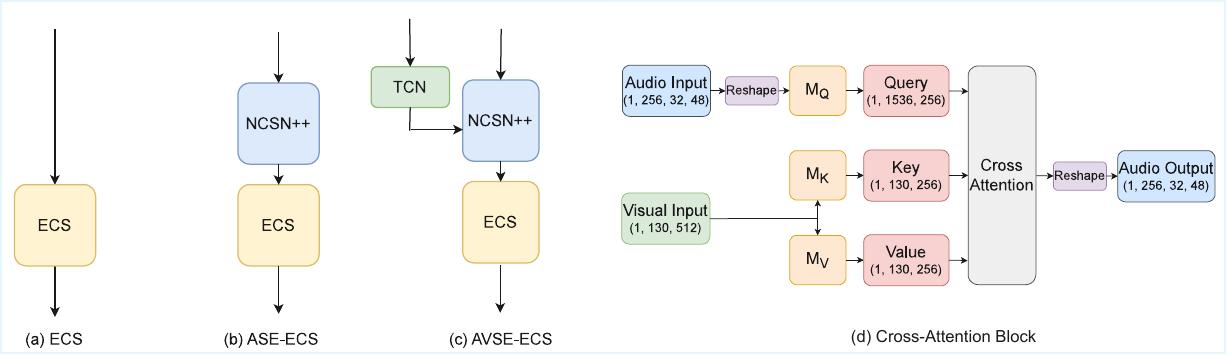}
    \caption{Neural network implementations: (a) ECS, (b) ASE-ECS, (c) AVSE-ECS, and (d) cross-attention block. As illustrated in Fig. \ref{fig:AVSE_ECS_framework}, the color blocks represent the modules of ECS (yellow), NCSN++ (blue), and TCN visual encoder (green).}
	%\vspace{-4mm}
    \label{fig:Arch_comp}
\end{figure}

This study explores the use of cross-attention blocks (red block in Fig. \ref{fig:AVSE_ECS_framework} (b)) to integrate audio and visual information. The attention mechanism enables neural networks to focus on the most relevant input parts, capture contextual dependencies, and facilitate alignment across modalities. Following a similar approach to \cite{DAVSE}, the self-attention blocks within the UNet architecture are modified by replacing the audio keys and values with visual embeddings generated by the visual encoder. Augmenting the audio input with lip movement information is intended to utilize visual cues to enhance speech reconstruction and thereby improve the overall performance of the SE system. To explain the design of the audio-visual fusion module, the cross-attention mechanism is shown in Fig. \ref{fig:Arch_comp} (d). Audio features (in blue) and visual features (in green) are provided as input, with dimensions indicated numerically. ${\bf M_Q}$, ${\bf M_K}$, and ${\bf M_V}$ represent the transformation layers for query, key, and value, respectively. They are fused by the cross attention block, then reshaped as audio output.

\subsection{Joint Training}

Joint training, as shown in Fig. \ref{fig:AVSE_ECS_framework} (b), simultaneously optimizes both the AVSE pre-processing module (blue block) and the ECS module (yellow block) within an end-to-end framework. To guide the SE module toward generating high-quality enhanced speech during the learning process, a training objective called spectrogram loss is defined as follows:
\begin{equation}
	\begin{array}{l}
		L_{{\rm Spec}} = {MSE} \left( {\bf E}, {\bf C} \right), 
	\end{array}
	\label{eq:Spec_loss}
\end{equation}
where \textit{MSE} is the mean squared error between ${\bf E}$, the enhanced speech from the AVSE model, and ${\bf C}$, the clean ground truth.
To measure the distance for each enhanced-clean electrodogram pair, the electrodogram loss is defined as
\begin{equation}
	\begin{array}{l}
		L_{{\rm Elec}} = {MSE} \left( {\bf \hat{E}}, {\bf \hat{C}} \right),
	\end{array}
	\label{eq:Elec_loss}
\end{equation}
where ${\bf \hat{E}}$ is the enhanced electrodogram from the AVSE-ECS model and ${\bf \hat{C}}$ is the clean electrodogram from the ECS model. The clean electrodogram is derived from the output of the ECS model when clean speech is provided as input.
By narrowing the distance between the clean electrodogram and the enhanced electrodogram, the output electrode pattern in AVSE-ECS can be further refined, reducing the impact of input noise and improving speech intelligibility.

By considering $L_{\rm Spec}$ and $L_{\rm Elec}$ concurrently, the total loss is defined as
\begin{equation}
	\begin{array}{l}
		L_{{\rm Total}} = \alpha * L_{{\rm Spec}} + \beta * L_{{\rm Elec}},
	\end{array}
	\label{eq:Total_loss}
\end{equation}
where ${\alpha}$ and ${\beta}$ are weights for $L_{\rm Spec}$ and $L_{\rm Elec}$, respectively. 

\section{Experiments}
\label{sec:experiments}

This section presents the experimental setups in this study.

\subsection{Objective Evaluation Metrics}
To assess the performance of CI systems, three commonly-used objective evaluation metrics are employed: short-time objective intelligibility (STOI) \cite{STOI}, extended short-time objective intelligibility (ESTOI) \cite{ESTOI}, and normalized covariance measure (NCM) \cite{NCM_2}. 
These metrics estimate speech intelligibility by measuring the correlation between CI-vocoded and clean speech. Furthermore, the neural models are compared using the signal-to-error ratio (SER) \cite{SER_2015}, where the error components include remaining noise after SE processing as well as artifacts and distortions introduced by the SE model.

\subsection{Dataset}
The audiovisual dataset used in this study is based on the Taiwan Mandarin Hearing in Noise Test (TMHINT) speech material \cite{MHINT_wong2007}, which consists of 16 phonemically balanced lists.
Each list contains 20 sentences, with each sentence consisting of 10 syllables (Chinese characters).
The duration of each utterance ranges from approximately 2 to 4 seconds. The audio-only dataset and the audio-visual dataset are both based on TMHINT sentences. The audio-only TMHINT dataset is used to train the ECS model, as described in \cite{ElectrodeNet_huang2024}, with the exception that 320 clean sentences from Lists 1 to 8 are used in this study.

For training and inference of the AVSE-ECS system, the Taiwan Mandarin Speech with Video (TMSV) dataset \cite{AVSE_light_chuang2022}, consisting of speech and video recordings of TMHINT sentences, is adopted. The method for using the TMSV dataset is similar to that described in \cite{AVSE_light_chuang2022}. This dataset includes speech recordings from 18 native Mandarin speakers (13 males and 5 females). In this study, considering gender balance, 8 speakers (4 male, 4 female) from TMSV are selected for training and validation. For each speaker with 320 utterances, sentences 1-190 are used for training, and sentences 191-200 for validation. These speech files are mixed with 100 types of noise \cite{Training_Noise} at five signal-to-noise ratio (SNR) levels (-12, -6, 0, 6, 12 dB). To reduce training time, a total of 12,000 utterances are randomly sampled. For evaluation, utterances 201–320 from two unseen speakers (1 male, 1 female) are used. 
These are corrupted with five noise types including music, street noise, engine noise, baby cry, and pink noise, and mixed at SNR levels of -1, -4, -7, and -10 dB, yielding a total of 4,800 noisy utterances.
The training and testing sets are fully mismatched in terms of speech content, noise types, speakers, and SNR levels to ensure robust generalization.

\subsection{Training Configuration}
According to the total loss $L_{\rm Total}$ introduced in Section 2.6, two types of training configurations are designed to train the AVSE-ECS Network: pre-trained and joint training. For the pre-trained method, the AVSE model is trained independently and concatenated with ECS. For the joint training method, an untrained AVSE network is connected to ECS, and the training of the full AVSE-ECS network includes the spectrogram loss $L_{\rm Spec}$ and the electrodogram loss $L_{\rm Elec}$. The pre-trained method can be regarded as a special case of joint training, which only includes $L_{\rm Spec}$ in the training stage by setting ${\beta}$ = 0. Note that in all configurations, the ECS models are pre-trained and concatenated after the AVSE model. During the joint training stage, all the parameters in ECS are frozen, and only the parameters in the AVSE model are tuned.

\subsection{Objective Evaluation}
The following three experiments are conducted for objective evaluation.
\begin{enumerate}
\itemsep=0pt
\item ECS with clean and noisy speech
\item AVSE-ECS with different $\beta$ values
\item Comparison of ACE, ECS, ASE-ECS, AVSE-ECS
\end{enumerate}

In the first experiment, the pre-trained ECS model is evaluated on the TMSV test set using STOI, ESTOI, and NCM metrics under both clean and noisy conditions. All evaluations are conducted on unseen speakers, with the objective scores for the noisy condition averaged over different noise types and SNR levels.

The second experiment examines how model weights in joint training affect speech intelligibility. Following the pre-trained method with $\alpha$ = 1, and different $\beta$ values are studied. The result helps select $\beta$ for the next experiment.

The third experiment compares three CI network implementations: ECS (Fig. \ref{fig:Arch_comp}(a)), ASE-ECS (Fig. \ref{fig:Arch_comp}(b)), and AVSE-ECS (Fig. \ref{fig:Arch_comp}(c)), with the common ACE coding strategy as the baseline. Apart from the ECS model without the SE function, the ASE-ECS model is an audio-only system in which self-attention replaces the cross-attention mechanism used in the AVSE-ECS model. AVSE-ECS, the proposed end-to-end CI system, is evaluated using both the pre-trained method and the joint training method. For all experiments, ${\alpha}$ is set to 1, and in the joint training setup, ${\beta}$ is set to 0.5. 

\section{Results and Discussion}
\label{sec:results}

\subsection{ECS with clean and noisy speech}
The overall objective intelligibility of the ECS model under clean and noisy conditions is shown in Table \ref{tab1}. For clean speech, ECS achieves an STOI score of 0.7604, which drops to 0.4870 under noisy conditions, averaged across speakers, noise types and SNR levels. Similar declines in ESTOI and NCM occur as conditions shift from clean to noisy. The ECS model’s performance declines under noisy conditions, as it only simulates the ACE strategy without SE capability.

\begin{table}
    \caption{Objective evaluation of the ECS model}
    \centering
    \renewcommand{\arraystretch}{0.6} 
    \begin{tabular}{cccc}
    \hline\hline
        Input Speech & STOI${\uparrow}$  & ESTOI${\uparrow}$ & NCM${\uparrow}$ \\
        \hline
        Clean & 0.7604 & 0.5866 & 0.6809 \\
        Noisy & 0.4870 & 0.2073 & 0.3258 \\
    \hline\hline
    \end{tabular}
    \label{tab1}
\end{table}
    
% Adjust this value to control the space
\vspace{-1em}

\begin{table}
    \caption{Objective evaluation of the AVSE-ECS network with different $\beta$ values under noisy conditions}
    \centering
    \renewcommand{\arraystretch}{0.6} 
    \begin{tabular}{cccc}
    \hline\hline
        $\beta$ & STOI${\uparrow}$  & ESTOI${\uparrow}$ & NCM${\uparrow}$ \\
        \hline
        1.0 & 0.6196 & 0.3705  & \textbf{0.5217} \\
        0.5 & \textbf{0.6305} & \textbf{0.3899} & 0.5211 \\
        0.25 & 0.6295 & 0.3879 & 0.5199 \\
    \hline\hline
    \end{tabular}
    \label{tab2}
\end{table}

\subsection{AVSE-ECS with different $\beta$ values}
Table \ref{tab2} presents the objective evaluation results of AVSE-ECS with different joint training setups. In noisy conditions, AVSE-ECS attains STOI scores of 0.6295 ($\beta$ = 0.25), 0.6305 ($\beta$ = 0.5), and 0.6196 ($\beta$ = 1), averaged over speakers, noise types and SNR levels. Similar trends in STOI and ESTOI scores appear across the three $\beta$ values. These results suggest that emphasizing $\beta$, the weight on the electrodogram loss $L_{\rm Elec}$, to 1 does not necessarily improve speech intelligibility. On the other hand, a smaller $\beta$ value of 0.25 limits the refinement of the electrodogram. Among the tested values, $\beta$ = 0.5 results in the highest STOI and ESTOI scores. Although $\beta = 1$ achieves a slightly higher NCM score (0.5217) compared to $\beta = 0.5$ (0.5211), the difference is considerably small. Therefore, the $\beta$ value of 0.5 is selected for the subsequent experiment.

\begin{table}
    \caption{Objective evaluation of various CI implementations under noisy conditions}
    \centering
    \renewcommand{\arraystretch}{0.6} 
    \begin{tabular}{lccccc}
    \hline\hline
        Method & Configuration & STOI${\uparrow}$  & ESTOI${\uparrow}$ & NCM${\uparrow}$ & $\Delta$SER (dB)${\uparrow}$ \\
        \hline
        ACE  & - & 0.4870 & 0.2067 & 0.3262 & - \\
        ECS  & - & 0.4870 & 0.2073 & 0.3258 & 0.0761\\
        ASE-ECS & Pre-trained & 0.5943 & 0.3557 & 0.4914 & 3.8075 \\
        AVSE-ECS  & Pre-trained & 0.6141 & 0.3672 & 0.5067 & 3.7291\\
        AVSE-ECS  & Joint training & \textbf{0.6305} & \textbf{0.3899} & \textbf{0.5211} & \textbf{7.4666} \\
    \hline\hline
    \end{tabular}
    \label{tab3}
\end{table}

\subsection{Comparison of ACE, ECS, ASE-ECS, and AVSE-ECS}
Table \ref{tab3} compares various CI system implementations under noisy conditions, with objective evaluation scores averaged across speakers, noise types and SNR levels. The ECS model achieves an average STOI score of 0.4870, comparable to the ACE strategy. With the pre-trained setup, the audio-only ASE-ECS model achieves a STOI score of 0.5943, surpassing the ECS strategy without the SE module. Incorporating visual cues in the AVSE-ECS model further increases STOI scores to 0.6141 (pre-trained) and 0.6305 (joint training). Similar trends appear in ESTOI and NCM scores. 

The average SER scores for all models are higher than that of the ACE strategy. The SER difference for the ECS model compared to ACE is 0.0761 dB, while the scores for ASE-ECS (3.8075 dB) and pre-trained AVSE-ECS (3.7291 dB) are close to each other. Moreover, the joint-training model achieves the highest SER score, exceeding ACE by 7.4666 dB.

Therefore, integrating ASE and AVSE models improves objective speech intelligibility over the ECS model, while the AVSE-ECS model with visual cues outperforms the audio-only ASE-ECS model in STOI, ESTOI, and NCM scores. In joint training that incorporates the clean electrodogram as an auxiliary target, the AVSE model can directly generate enhanced electrode stimulation patterns and achieve the highest scores in this study.

\subsection{Discussion}
This study can be further extended. Beyond the comprehensive objective evaluations, subjective listening tests are planned for the proposed AVSE-ECS method with both normal-hearing participants using CI simulations and actual CI users. Additionally, physiological evaluations, such as responses to auditory stimuli processed by the proposed approaches, may provide further objective evidence in human subjects.

In this study, the objective evaluation relies on a single TMSV dataset in one language, Mandarin Chinese. As highlighted in prior studies \cite{TTS_languate_zhang2019, TTS_languate_sankar2024, Mamba_chao2025}, regression-based tasks such as speech enhancement and speech separation are generally less sensitive to language differences. As a result, models trained on English speech data have shown effective generalization to other languages.
Moreover, to ensure the cross-language, cross-speaker, cross-dataset, and cross-validation generalizability of the proposed approach, further studies using diverse datasets are planned.
Analyses with a wider range of noise types are to be conducted, as the present results indicate no clear differences among the five noise types examined.
Since audiovisual datasets are less accessible than audio-only ones, further dataset collection and development are needed in terms of size, diversity, and real-world coverage.
For each dataset, multiple runs with different random seeds will be repeated to ensure the consistency and robustness of the proposed system.

Issues in implementing AVSE in CI hardware need to be tackled.
The proposed AVSE system with end-to-end CI processing can be executed on a personal computer, making it suitable for applications with relaxed timing constraints, such as movie streaming, online courses, and playback scenarios. Latency is challenging for real-time deployment on CI sound processors, but integration with external edge devices such as smartphones, tablets, or smart glasses can offload computation from implant-side processors, enabling more advanced audiovisual processing in everyday use. Looking ahead, we plan to reduce the model’s computational load through techniques such as pruning and knowledge distillation. Further investigations into the NCSN++ model, the proposed AVSE backbone, will assess lightweight alternatives for hardware implementation and seamless integration into the hearing assistance ecosystem. In the future, advances in device processing power are expected to reduce signal processing latency and make the AVSE module a feasible application.

It is worth noting to consider a number of related topics. A recent study demonstrates an analogous architecture to an end-to-end neural amplifier designed for hearing aids \cite{NeuroAMP}. Therefore, more advanced AVSE pre-processing methods and end-to-end frameworks will be explored to enhance the overall CI sound processing system. Furthermore, the role of multisensory cues in CI perception will be investigated to provide more natural sound and enhance communication and well-being for CI users. Research on deep learning and CI sound processing is an ongoing journey.

\section{Conclusion}
\label{sec:conclusion}

This paper introduces a novel end-to-end CI system, AVSE-ECS, which integrates an audio-visual speech enhancement (AVSE) model with a deep-learning-based sound coding strategy, ElectrodeNet-CS (ECS). This system exhibits strong noise suppression capabilities and effectively mitigates severe performance degradation of the CI in noisy environments. Joint training of the end-to-end system enhances electrode stimulation patterns for greater clarity and recognizability, as verified through experiments. Future work includes subjective listening evaluations, cross-dataset generalization studies, and exploration of alternative AVSE preprocessing frameworks. The methods and findings of this study may provide insights into end-to-end CI coding strategies, multimodal processing, and related fields.

\section{Author Declarations}
\subsection{Conflict of Interest}
The authors have no conflicts to disclose.

\subsection{Ethics Approval}
This study was based solely on CI simulations and did not involve human participants, patient data, or animal subjects.

\section{Data Availability}
The data supporting the results of this research are available from the corresponding author upon reasonable request.

\bibliography{references}

\end{document}